\pdfoutput=1
\documentclass[11pt,twoside,a4paper,cmspaper,final,collab]{cms-tdr}
\def\svnVersion{c7852ff-D}\def\svnDate{2021/05/18}\def\cmsCernNoTag{CERN-EP-2021-020}\def\cmsCernDate{\today}\def\cmsMessage{"Published in Physical Review Letters as \href{http://dx.doi.org/10.1103/PhysRevLett.126.252003}{\doi{10.1103/PhysRevLett.126.252003}.}"}
\begin{document}\cmsNoteHeader{BPH-20-004}

\newcommand{\mumu}      {\ensuremath{\PGmp\PGmm}\xspace}
\newcommand{\pins}      {\PGp}
\newcommand{\pip}       {\PGpp}
\newcommand{\pim}       {\PGpm}
\newcommand{\pipm}      {\PGppm}

\newcommand{\Km}        {\PKm}

\newcommand{\jpsi}      {\PJGy}

\newcommand{\La}        {\PGL}
\newcommand{\Xim}       {\PGXm}

\newcommand{\Xb}        {\PGXmb}
\newcommand{\Xbz}       {\PGXzb}
\newcommand{\Xbns}      {\PGXb}

\newcommand{\Xbs}       {\PGXstzb}
\newcommand{\Xbsns}     {\PGXstb}

\newcommand{\Xbst}      {\ensuremath{\PGXb^{\!\!\!\ast{}\ast{}}}\xspace}
\newcommand{\Xbstm}     {\ensuremath{\PGXb^{\!\!\!\ast{}\ast{}-}}\xspace}

\newcommand{\Xbpz}      {\PGXprzb}
\newcommand{\Xbpns}     {\PGXprb}

\newcommand{\Xbss}      {\ensuremath{\PGXb\hspace{-.09em}(6100)^-}\xspace}
\newcommand{\bsq}       {\ensuremath{\PQb\PQs\PQq}\xspace}
\newcommand{\qs}        {\ensuremath{\PQq\PQs}\xspace}
\newcommand{\PVPV}      {\ensuremath{\text{PV}}\xspace}

\newcommand{\Lb}        {\PGLzb}
\newcommand{\XbJX}      {\ensuremath{\Xb\to\jpsi\Xim}\xspace}
\newcommand{\XbJSK}     {\ensuremath{\Xb\to\jpsi\PGSz\Km}\xspace}

\newcommand{\XbJLK}     {\ensuremath{\Xb\to\jpsi\La\Km}\xspace}

\newcommand{\JpsiXim}   {\ensuremath{\jpsi\Xim}\xspace}
\newcommand{\JpsiLaKm}  {\ensuremath{\jpsi\La\Km}\xspace}
\newcommand{\XbP}       {\ensuremath{\Xb\pip}\xspace}
\newcommand{\XbPP}      {\ensuremath{\Xb\pip\pim}\xspace}

\newcommand{\XiLamPi}   {\ensuremath{\Xim\to\La\pim}\xspace}
\newcommand{\LamPPi}    {\ensuremath{\La\to\Pp\pim}\xspace}
\newcommand{\SiLaGa}    {\ensuremath{\PGSz\to\La\PGg}\xspace}

\newcommand{\Xbsd}      {\ensuremath{\Xbs\to\XbP}\xspace}
\newcommand{\Xbssd}     {\ensuremath{\Xbss\to\XbPP}\xspace}
\newcommand{\Xbssdi}    {\ensuremath{\Xbss\to\Xbs\pim}\xspace}

\newcommand{\dmX}       {\ensuremath{\Delta M}\xspace}
\newcommand{\dmXm}      {\ensuremath{\Delta M_{\Xbss}}\xspace}

\newcommand{\chisq}     {\ensuremath{\chi^2}\xspace}
\newcommand{\ips}       {\ensuremath{d_{xy}/\sigma_{d_{xy} }}\xspace}
\newcommand{\pvtx}      {\ensuremath{P_{\text{vtx}}}\xspace}

\newcommand{\sq}[1]   {\ensuremath{\sqrt{s} = #1}\xspace}

\newcommand{\pp}        {\ensuremath{\Pp\Pp}\xspace}

\ifthenelse{\boolean{cms@external}}{\providecommand{\cmsLeft}{upper\xspace}}{\providecommand{\cmsLeft}{left\xspace}}
\ifthenelse{\boolean{cms@external}}{\providecommand{\cmsRight}{lower\xspace}}{\providecommand{\cmsRight}{right\xspace}}
\newlength\cmsFigWidth\ifthenelse{\boolean{cms@external}}{\setlength\cmsFigWidth{0.48\textwidth}}{\setlength\cmsFigWidth{0.75\textwidth}}

\cmsNoteHeader{BPH-20-004}
\title{Observation of a new excited beauty strange baryon decaying to \texorpdfstring{\XbPP}{Xib pi+ pi-}}
\date{\today}

\abstract{
    The \XbPP invariant mass spectrum is investigated
    with an event sample of proton-proton collisions at $\sq{13}\TeV$, collected by the CMS experiment at the LHC 
    in 2016--2018 and corresponding to an integrated luminosity of 140\fbinv.
    The ground state \Xb is reconstructed via its decays to \JpsiXim and \JpsiLaKm.
    A narrow resonance, labeled \Xbss, is observed at a \XbPP invariant mass of  
    $6100.3 \pm 0.2\stat\pm 0.1\syst \pm 0.6\,(\Xb)\MeV$,
    where the last uncertainty reflects the precision of the \Xb baryon mass.
    The upper limit on the \Xbss natural width is determined to be 1.9\MeV at 95\% confidence level. 
    The low \Xbss signal yield observed in data does not allow
    a measurement of the quantum numbers of the new state.
    However, following analogies with the established excited \PGXc baryon states, 
    the new \Xbss resonance and its decay sequence are consistent
    with the orbitally excited \Xb baryon, with spin and parity quantum numbers $J^P=3/2^-$.   
}

\hypersetup{
    pdfauthor={CMS Collaboration},
    pdftitle={Observation of a new excited beauty strange baryon decaying to Xib- pi+ pi-},
    pdfsubject={CMS},
    pdfkeywords={CMS, beauty baryon, heavy baryon, flavour spectroscopy}}

\maketitle 

The \Xbns baryon family consists of isodoublet states composed of \bsq quarks,
where \PQq represents an up or a down quark for the \Xbz and \Xb states, respectively. 
According to the quark model for baryons containing one heavy quark~\cite{Th:diquark},
three such isodoublets that are neither orbitally nor radially excited
should exist, including one with the light diquark angular momentum $j_{\qs}=0$ 
and spin-parity $J^P = 1/2^+$ (the \Xbns ground states), 
one with $j_{\qs}=1$ and $J^P=1/2^+$ (the \Xbpns),
and one with $j_{\qs}=1$ and $J^P=3/2^+$ (the \Xbsns). 
Various theoretical models and calculations predict a spectrum 
of excited \Xbns baryons~\cite{Th:m_bb_rel_quark_model, Th:b_bar_1/Nc, Th:b_baryons, Th:bb_in_quark_model, Th:regge_bb_rel, Th:faddeev_bb_spectr, Th:LQ_XQ_heavy_quark, Th:ITEP_Trusov, Th:sum_rule_Pwave, Th:12_32_heavy, Th:S_P_waves_single_bb, Th:heavy_w_chiral, Th:bb_hyperfine, Th:regge_traj, Th:spectr_123_bb}.
Three of the four excited states with $j_{\qs} = 1$ have been observed 
at the CERN LHC~\cite{BPH12001, LHCb_XibStar0, LHCB_XibPrime-Star-} via their
\XbP and $\Xbz\pim$ decays, in agreement with 
predictions~\cite{Th:b_bar_1/Nc, Th:b_baryons, Th:m_bb_rel_quark_model}. 
The fourth state, \Xbpz, is expected to be lighter than the \XbP mass threshold,
making a strong transition to \Xb kinematically impossible. 
The next prominent isodoublets, in analogy with the quark model assumptions for
the well-established excited \PGXc baryons~\cite{PDG}, 
are orbitally excited $P$-wave \Xbst states with $J^P=1/2^-$ ($3/2^-$),
expected to decay to $\Xbpns(\Xbsns)\pins$~\cite{Th:S_P_waves_single_bb, Th:heavy_w_chiral, Th:Xb(6227)_impact}.
Recently, the LHCb Collaboration reported the observation of the $\Xbns(6227)^-$~\cite{LHCb_Xib6227}
and $\Xbns(6227)^0$~\cite{LHCb_Xib62270} states,
the former decaying to both $\Lb\Km$ and $\Xbz\pim$, 
and the latter to \XbP. 

\begin{figure}[h]
    \centering
    \includegraphics[width=\cmsFigWidth]{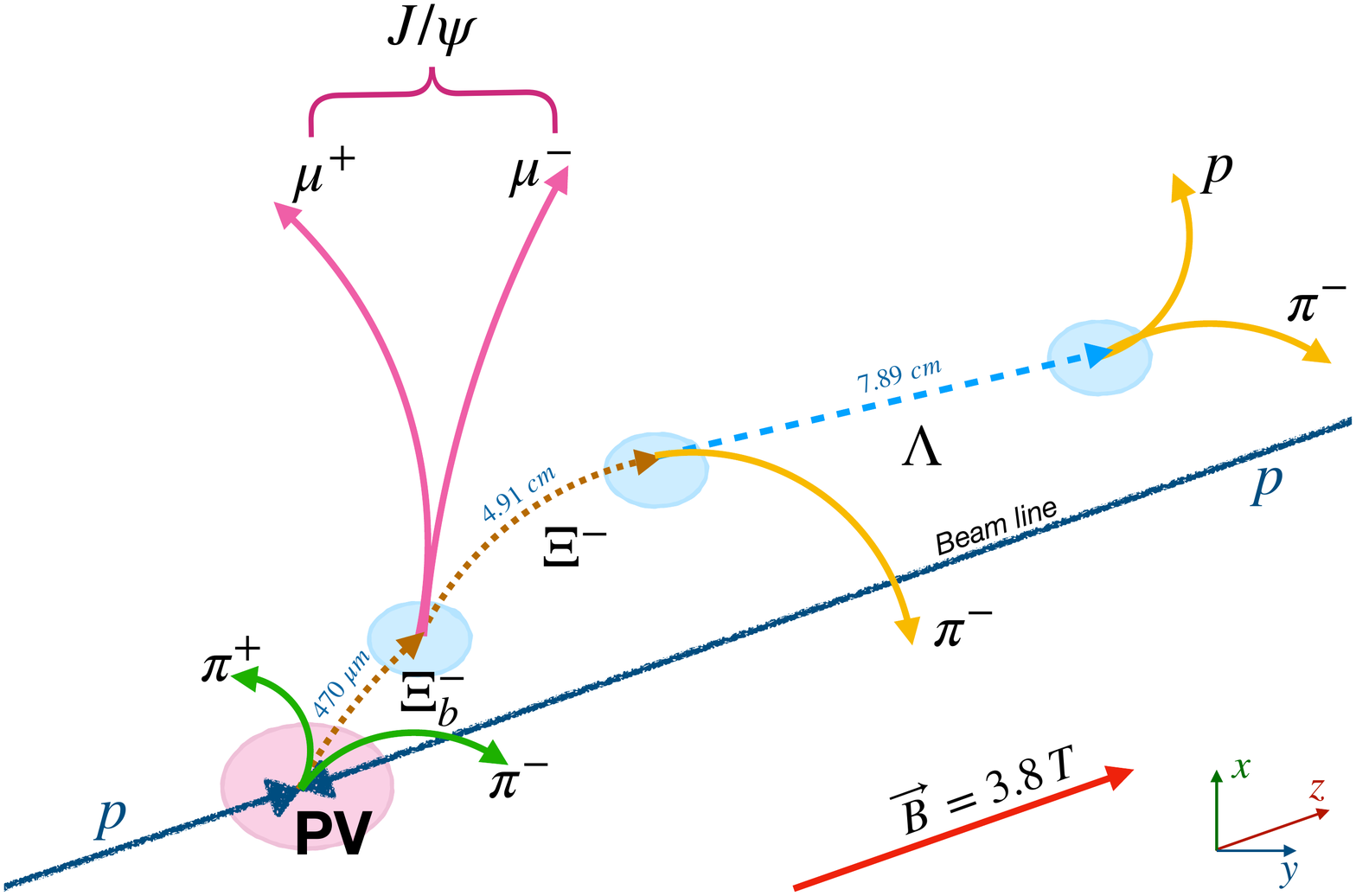}
    \includegraphics[width=\cmsFigWidth]{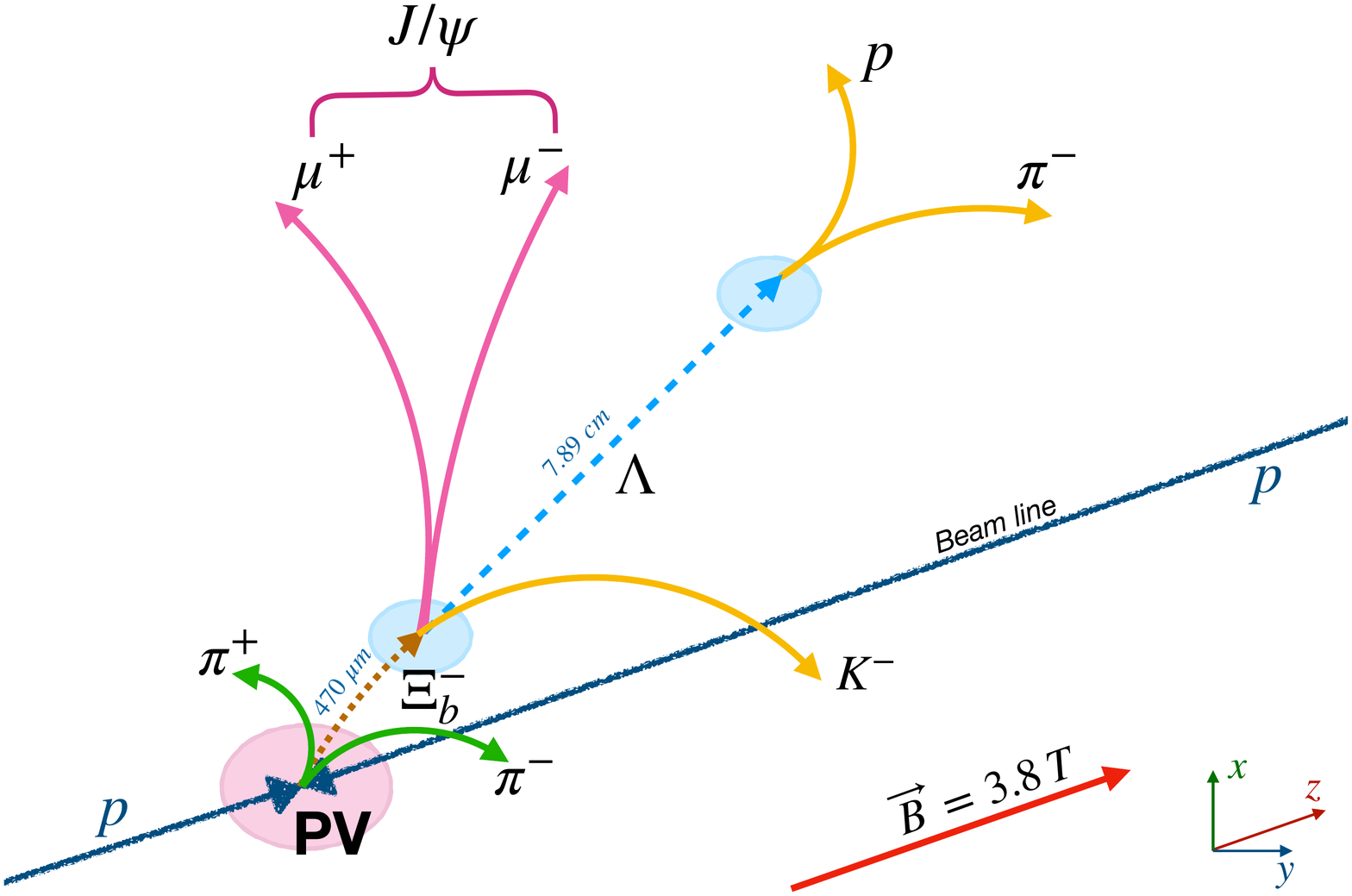}
    \caption{
        The $\Xbss\to\PGXmb\PGpp\PGpm$ decay topology,
        where the \PGXmb decays to $\PJGy\PGXm$ (\cmsLeft) or to $\PJGy\PGL\PKm$ (\cmsRight).
        The numbers in blue are average decay lengths.
    } 
    \label{fig:decay_topo}
\end{figure}

This Letter presents a search for \Xb excited states in the \XbPP invariant mass spectrum, 
performed using proton-proton (\pp) collision data samples collected by the CMS experiment
at the LHC at $\sq{13}\TeV$ in 2016--2018, corresponding to an integrated luminosity
of 140\fbinv. The ground state \Xb is reconstructed via its decays
to \JpsiXim and \JpsiLaKm, followed by the decays $\jpsi\to\mumu$, $\XiLamPi$, and $\LamPPi$. 
The decay topologies are illustrated in Fig.~\ref{fig:decay_topo}. 
For the \XbJLK decay mode, following the studies reported by the LHCb Collaboration~\cite{LHCb_JpsiLamK}, 
the partially reconstructed \XbJSK channel 
is also used, where the photon from the \SiLaGa decay is too soft to be detected.
The inclusion of charge-conjugated states is implied throughout this Letter.
A signal peak, hereafter referred to as \Xbss, is clearly observed
near the \XbPP kinematic threshold, with a decay sequence consistent with being the
$\Xbss\to\Xbs\pim\to\XbPP$ decay. The \Xbss mass and an upper limit on its width are also measured.

The central feature of the CMS apparatus is a superconducting solenoid of 6\unit{m} 
internal diameter, providing a magnetic field of 3.8\unit{T}. Within the solenoid 
volume are a silicon pixel and strip tracker, a lead tungstate crystal electromagnetic 
calorimeter, and a brass and scintillator hadron calorimeter, each composed of 
a barrel and two endcap sections.
Muons are detected in gas-ionization chambers embedded 
in the steel flux-return yoke outside the solenoid.
A more detailed description
of the CMS detector, together with a definition of the coordinate system used and
the relevant kinematic variables, can be found in Ref.~\cite{CMS}.

Events of interest are selected using a two-tiered trigger system.
The first level (L1), composed of custom hardware processors, 
uses information from the calorimeters and muon detectors~\cite{Sirunyan:2020zal}. 
The second level, known as the high-level trigger (HLT), consists of a farm of processors 
running a version of the full event reconstruction software
optimized for fast processing~\cite{Khachatryan:2016bia}. 
The events used in the analysis were selected at L1 by requiring 
the presence of at least two muons, and at HLT by requiring 
that the two muons have opposite sign (OS), 
with various thresholds on the pseudorapidity $\eta$ and momentum transverse to the beam axis \pt,
compatible with being produced in the dimuon decay of \jpsi mesons.

Several simulated event samples are used in the analysis.
The \PYTHIA~8.230 package~\cite{Pythia8} is used to simulate the production of the \Xbss state,
where the \PGSmb baryon, with a modified mass value, is used as a proxy for an excited \Xbss state.
The \Xbssd (including both resonant $\Xbs\pip\to\XbPP$ and non-resonant \XbPP modes), 
\XbJX, \XbJLK (including \XbJSK, \SiLaGa), and $\jpsi\to\mumu$ decays
are modeled with \EVTGEN 1.6.0~\cite{EvtGen}, where final-state photon 
radiation is included using \PHOTOS 3.61~\cite{PHOTOS,PHOTOS_2}.
The generated events are then passed to a detailed \GEANTfour-based
simulation~\cite{GEANT4} of the CMS detector, including the
same trigger and reconstruction algorithms as used for the collision data.
The simulation includes effects from multiple \pp interactions in the same
or nearby bunch crossings (pileup) with a multiplicity distribution
matching the measured one.

The selection criteria are optimized using the Punzi figure of merit~\cite{Punzi:2003bu}, 
which does not rely on the signal normalization. 
The expected background is estimated from data using the same-sign (SS) control region
described below, while the signal efficiency is obtained from the simulated \Xbssd events. 
The \XbJX and \XbJLK requirements are optimized separately.

Events are required to have two OS muons passing the CMS soft-muon selection criteria~\cite{CMS_muons}
and satisfying $\pt(\PGmpm)>3\GeV$ and $\abs{\eta(\PGmpm)}<2.4$.   
The muons must form a common vertex with \chisq probability \pvtx above 1\%. 
The dimuon invariant mass must be 
within 100\MeV of $m_{\jpsi}^{\mathrm{PDG}}$ 
(hereafter, $m_{X}^{\mathrm{PDG}}$ denotes the world-average mass of hadron $X$~\cite{PDG}), 
corresponding to about three times the mass resolution.
The \La candidates are formed from displaced two-prong vertices, assuming the decay \LamPPi,
as described in Ref.~\cite{CMS_tracks_V0}.  
The $\Pp\pim$ reconstructed mass is required to be
within 10\MeV of $m_{\La}^{\mathrm{PDG}}$,
corresponding to about three times the mass resolution.
The two tracks are then refitted with
their invariant mass constrained to $m_{\La}^{\mathrm{PDG}}$.
The obtained \La candidates are required to have $\pt>1\GeV$ and $\pvtx>1\%$.

For the \XbJX channel, the \XiLamPi candidates are obtained by combining charged particles
of $\pt>0.25\GeV$ with the selected \La candidates. 
The reconstructed \Xim must have $\pvtx>1\%$, $\pt>3\GeV$,
and invariant mass within 9.5\MeV of $m_{\Xim}^{\mathrm{PDG}}$,
corresponding to about three times the mass resolution. 
The \Xb candidates are obtained by performing a $\mumu\Xim$ kinematic vertex fit, 
constraining the dimuon invariant mass 
to $m_{\jpsi}^{\mathrm{PDG}}$. 

{\tolerance=800
    For the \XbJLK decay channel, the \La candidates must pass stricter requirements:
    \mbox{$\pt>2\GeV$} and $\abs{M(\Pp\pim) - m^{\mathrm{PDG}}_\La}<9\MeV$.
    The charged kaon candidates are particle tracks with kaon mass assignment 
    satisfying high purity tracking requirements~\cite{CMS_tracks} and $\pt>1.2\GeV$.
    The \Xb candidates are reconstructed by fitting the $\mumu\La\Km$ 
    vertex with the \jpsi mass constraint. 
    Since the photon from the \SiLaGa decay is not detected, 
    both \XbJLK and \XbJSK decays contribute 
    to the $\mumu\La\Km$ reconstructed combination. 
    \par}

The \Xb candidates are required to have $\pvtx>1\%$ 
and $\pt>10\,(15)\GeV$ for the \XbJX (\XbJLK) channel.
From all reconstructed \pp collision vertices, the primary vertex (PV) 
is chosen as the one with the smallest pointing angle, as done in 
Refs.~\cite{BPH16002, BPH16003, BPH18007, BPH19003}.
The pointing angle is the three-dimensional angle between the \Xb candidate momentum and 
the vector joining the PV with the reconstructed \Xb candidate decay vertex.
The decay length $L_{xy}$ of the \Xb candidate in the transverse plane,
computed as the two-dimensional distance between the PV and the \Xb decay vertex,
is required to be at least three times larger than its uncertainty $\sigma_{L_{xy}}$. 
The $\vec{\pt}(\Xb)$ is required to be aligned with the
transverse displacement vector:
$\cos(\alpha(\Xb, \PVPV)) > 0.99\,(0.993)$ for the \XbJX (\XbJLK) channel,
where $\alpha(\Xb, \PVPV)$ is the pointing angle in the plane transverse to the beams.
Two additional topological requirements are applied: 
the cosine of the pointing angle, $\cos(\alpha(\Xim,\,\Xb))$, 
must be larger than 0.999 for the \XbJX channel;
and $L_{xy}/\sigma_{L_{xy}}(\La, \Xb) > 20$ for the \XbJLK channel. 
Besides, the pion emitted in the \XiLamPi decay and the kaon emitted in the \XbJLK decay
must have $\ips>0.9$ and 0.6,
respectively, where $d_{xy}$ is the impact parameter in the transverse plane with respect to the PV,
and $\sigma_{d_{xy}}$ is its uncertainty.

\begin{figure}[h!t]
    \centering
    \includegraphics[width=0.48\textwidth]{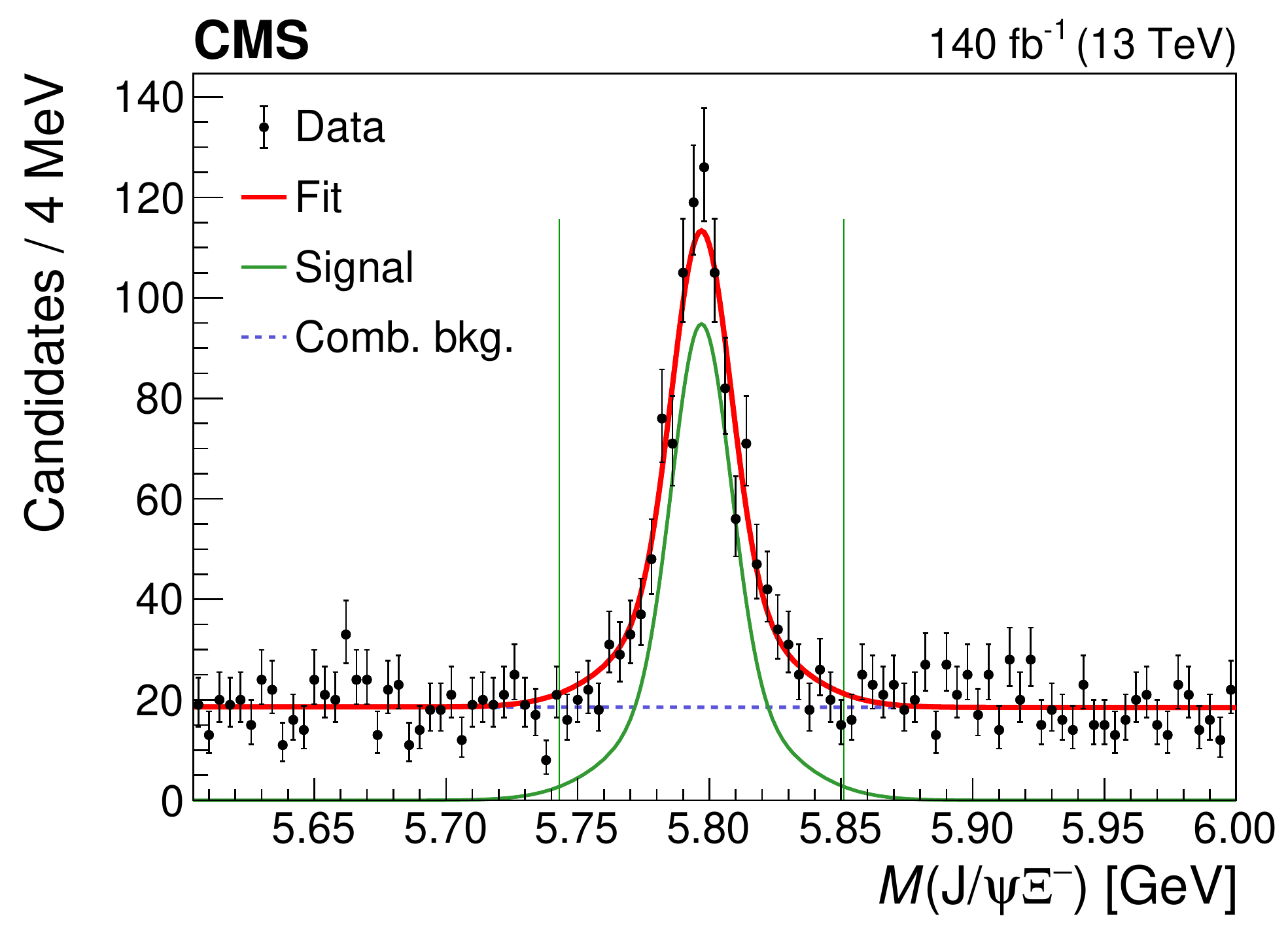}
    \includegraphics[width=0.48\textwidth]{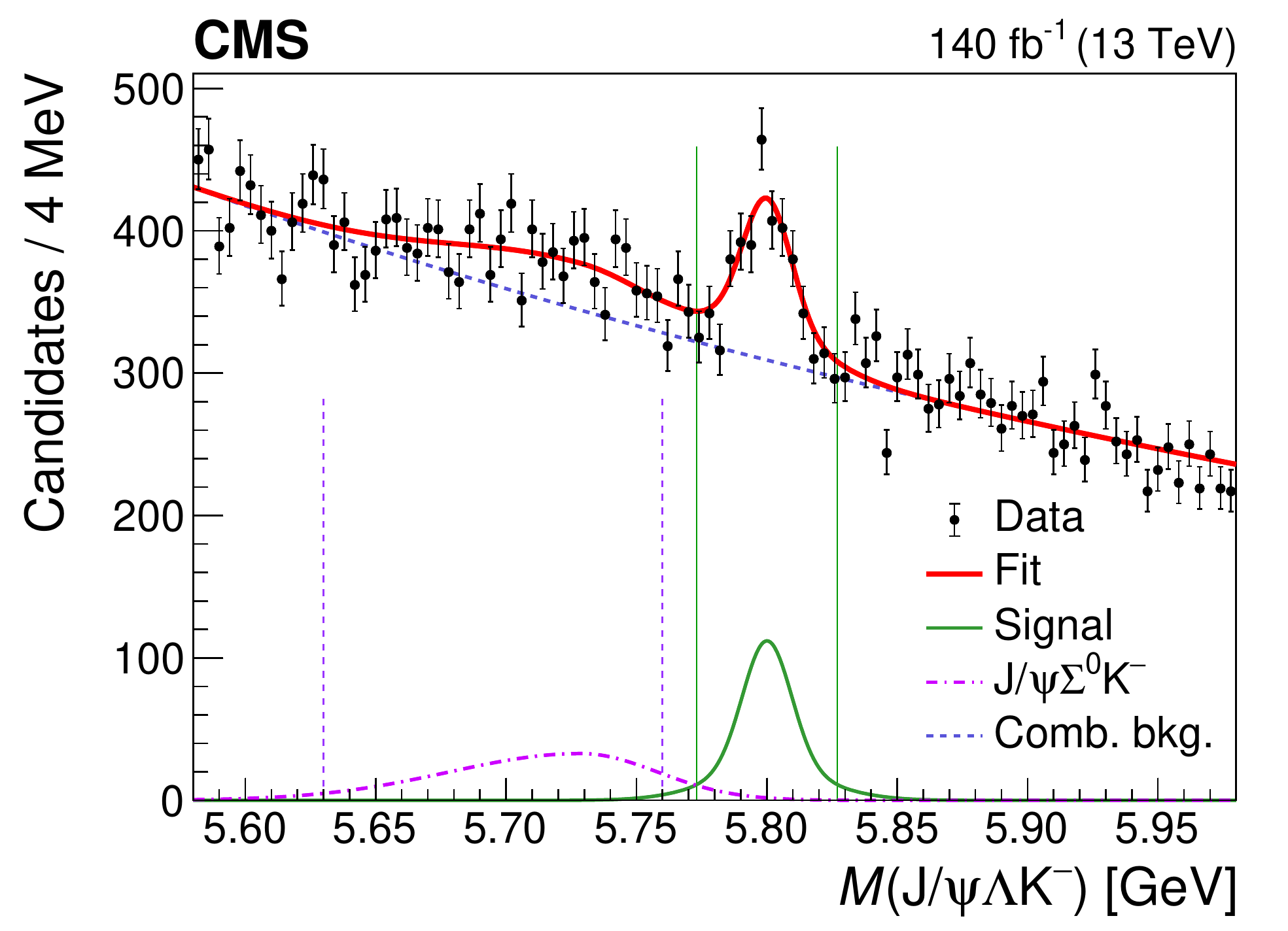}
    \caption{
        Invariant mass distributions of the selected \Xb candidates in the \JpsiXim (\cmsLeft)
        and \JpsiLaKm (\cmsRight) decay channels 
        with the fit results superimposed.
        The vertical solid (dashed) lines show the mass windows discussed in the text and used 
        in the reconstruction of the \XbPP candidates in \JpsiXim and \JpsiLaKm
        ($\jpsi\PGSz\Km$) channels. 
        \label{fig:Xb_signal}
    } 
\end{figure}

The invariant mass distributions of the selected \Xb candidates are shown in 
Fig.~\ref{fig:Xb_signal} for the \JpsiXim (\cmsLeft) and \JpsiLaKm (\cmsRight) channels.
The two plots also show the results of independent
unbinned extended maximum-likelihood fits. 
In both cases, the fully reconstructed \Xb signal 
is described by a double-Gaussian function with two free parameters: 
the common mean and the total yield; 
the two width parameters and the proportion of each Gaussian
are fixed from simulation studies. 
The background is described by a first-order polynomial in the \JpsiXim fit
and an exponential function in the \JpsiLaKm fit.
In the latter fit, the signal contribution from the partially reconstructed \XbJSK decays 
is taken into account by including an asymmetric Gaussian in the fit model,
with the shape parameters fixed from simulation studies.
All normalization values (signals and backgrounds) are free parameters of the fit. 

The signal yields from the fits described above are $859\pm36$ 
and $815\pm74$ for the \XbJX and fully reconstructed \XbJLK decay modes, respectively,
with the uncertainties being statistical only.
The fitted \Xb masses of $5797.0\pm0.7$ and $5800.1\pm1.2\MeV$,
respectively for the \JpsiXim and \JpsiLaKm channels, the uncertainties being statistical only,
are consistent with each other and with the world-average value, $5797.0\pm0.6\MeV$~\cite{PDG}.
The signal components corresponding to fully reconstructed \Xb candidates 
are shown by the solid green curves. 
The fitted yield of the partially reconstructed \XbJSK contribution, 
reconstructed as \JpsiLaKm, is $820\pm158$, 
represented by the dotted-dashed curve in Fig.~\ref{fig:Xb_signal} (\cmsRight).
The \Xb fit results illustrate this part of the reconstruction procedure and provide
the first confirmation of the \XbJLK decay observed by LHCb~\cite{LHCb_JpsiLamK}.

When reconstructing \XbPP candidates, we select events with \Xb invariant mass
within 54 (27)\MeV of the fitted \Xb mass
for the \JpsiXim (\JpsiLaKm) channel, corresponding 
to approximately 2.8 (1.8) times the mass resolution, 
as shown by the vertical solid lines in Fig.~\ref{fig:Xb_signal}.
The $5.63<M(\JpsiLaKm)<5.76\GeV$ mass region is used 
for the partially reconstructed \XbJSK decay mode,
shown by the dashed vertical lines in Fig.~\ref{fig:Xb_signal} (\cmsRight).
These mass ranges are selected through the same optimization procedure 
as used for the other selection criteria.

Since the lifetime of the excited \Xbns states is expected to be negligible, 
the \XbPP candidates are formed by combining the selected \Xb candidates with
two OS tracks originating from the PV, as in
Refs.~\cite{BPH16002, BPH16003, BPH18007, BPH19003}.
Combinations of a \Xb candidate with two SS pions from the PV are used as
a control channel and form the SS control region.
The analysis is performed using the 
mass difference variable $\dmX = M(\XbPP) - M(\Xb) - 2\,m^{\mathrm{PDG}}_\pipm$,
which has a better mass resolution than $M(\XbPP)$,
where $M(\Xb)$ represents the reconstructed \Xb mass.
According to the simulation studies, this variable also has the advantage of 
being insensitive to a potential mass shift caused by the fact that the photon emitted in the
\XbJSK, \SiLaGa decay sequence is not reconstructed.
Following the technique developed in Ref.~\cite{BPH19003}, the selected 
\Xb candidate and all tracks forming the PV are refit to a common vertex,
further improving the \XbPP invariant mass resolution of the fully reconstructed channels
from $1.39\pm0.11$ to $0.94\pm0.06\MeV$ (statistical uncertainties only), 
as obtained from simulation studies.

Theoretical studies~\cite{Th:S_P_waves_single_bb, Th:heavy_w_chiral, Th:Xb(6227)_impact}
and analogous decays of excited charm baryons~\cite{PDG, CLEO_Xic(2815)} suggest 
that the decay $\Xbstm\to\XbPP$ should proceed predominantly
through $\Xbstm\to\Xbs\pim$, followed by \Xbsd.
Therefore, an additional requirement is applied to enhance this contribution.  
As the \Xbs state has a mass of $5952.3 \pm 0.6\MeV$, 
the mass difference $M(\Xbs)-M(\Xb)-m^{\mathrm{PDG}}_\pip$ will peak at 15.73\MeV~\cite{PDG}.  
To avoid complications in understanding the \XbPP threshold,
we do not apply a minimum cut on this mass difference but simply require it
to be less than 20.73\MeV, with the 5\MeV addition found to be optimal 
when considering the \Xbs natural width and our detector resolution.

\begin{figure}[ht!]
    \centering 
        \includegraphics[width=0.48\textwidth]{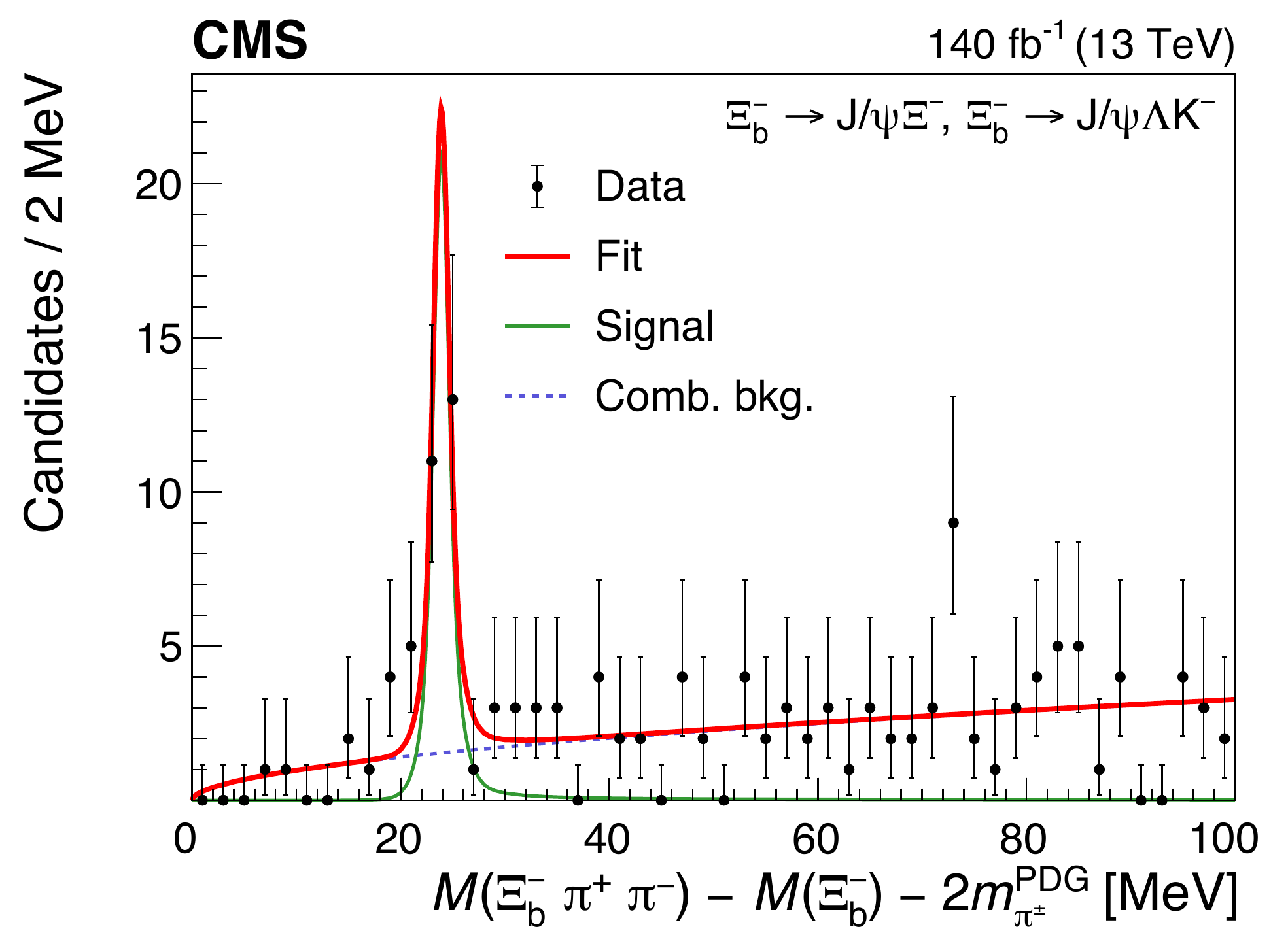}
        \includegraphics[width=0.48\textwidth]{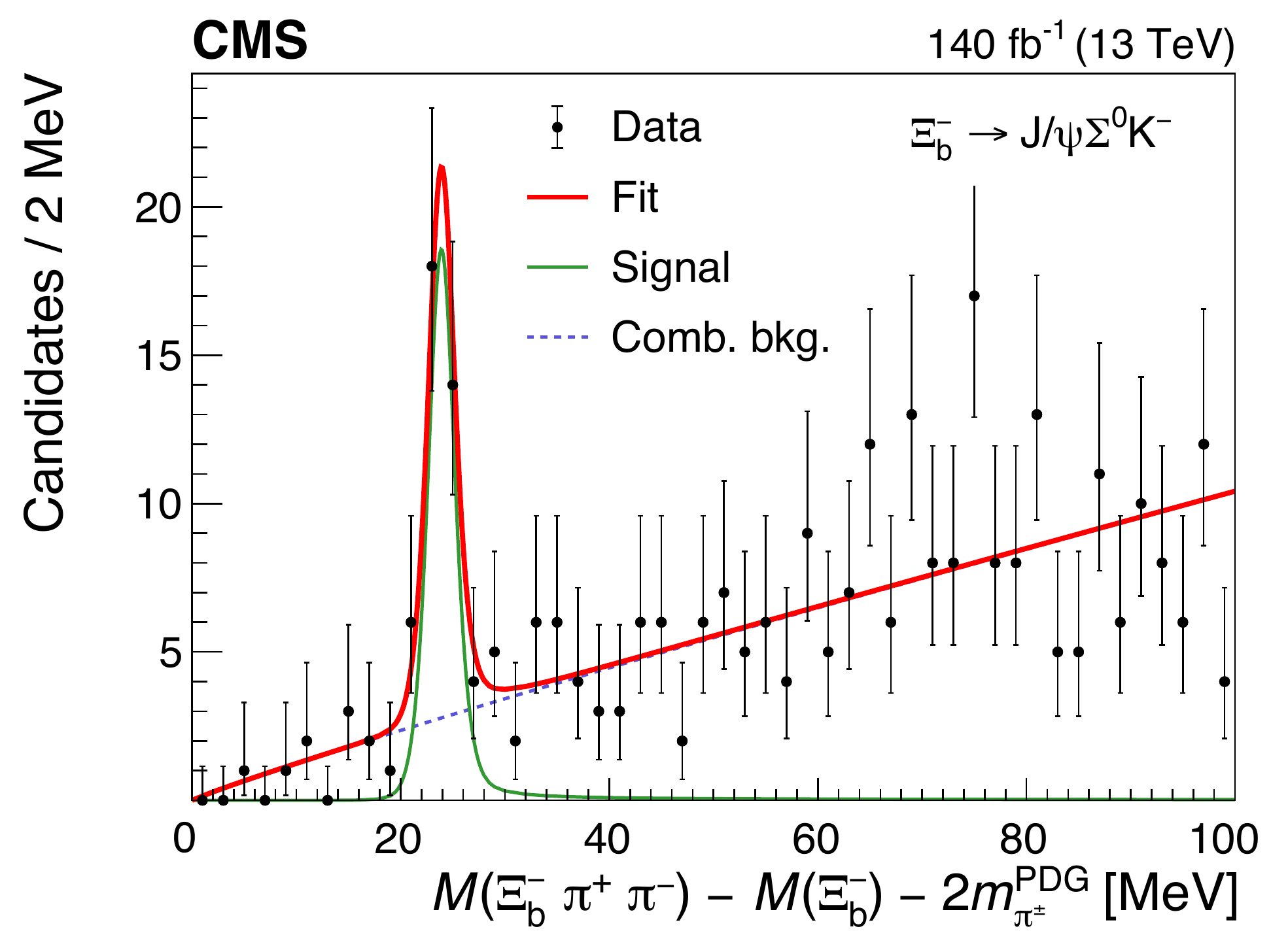}
    \caption{
        Distributions of the invariant mass difference \dmX for the selected \XbPP candidates,
        with the \Xb reconstructed in the \JpsiXim and \JpsiLaKm channels (\cmsLeft) or 
        partially reconstructed in the $\jpsi\PGSz\Km$ channel (\cmsRight).
        The result of the simultaneous fit is also shown.
        \label{fig:Xbss_comb}
    } 
\end{figure}

The invariant mass distribution of the selected \XbPP candidates 
is shown in Fig.~\ref{fig:Xbss_comb}, using the mass difference variable \dmX.
The \cmsLeft plot combines the data from the \XbJX and \XbJLK channels,
which have identical mass resolutions, according to simulation studies
(the \Xb is fully reconstructed in both channels).
The \cmsRight plot shows the events that use
the partially reconstructed \XbJSK channel, with a 30\% larger mass resolution.
Given the definition of the \dmX variable, the mean mass of the signal peaks should not depend on 
the \Xb reconstruction channel.

A narrow peak is seen near the threshold of the \XbPP system
in both plots of Fig.~\ref{fig:Xbss_comb}. 
The excess is also visible in each of the two independent decay channels, \JpsiXim and \JpsiLaKm.
We have also studied the OS and SS distributions in a wider range of \dmX 
(up to 280\MeV) and found no other significant peaks.
A simultaneous unbinned extended maximum-likelihood fit is performed on the 
two data samples shown in Fig.~\ref{fig:Xbss_comb}, the result being represented by the red curves.
The signal component is described with a relativistic 
Breit--Wigner (RBW) function~\cite{Jackson:1964zd,Blatt} for the \Xbssdi decay, 
convolved with a double-Gaussian resolution function.
The mass and natural width of the signal function are the two parameters of interest in the fit.
The normalization and background parameters are different 
for the fully and partially reconstructed channels, as are the resolution parameters, 
which are fixed from the simulation studies.
The background component is modeled with the threshold function 
$(\dmX)^\alpha$, where $\alpha$ is a free parameter.

The fitted mass difference of the new \Xbss state is $\dmXm=24.14\pm0.22\MeV$,
where the uncertainty is statistical only.
The fitted signal yields are $26\pm7$ and $34\pm9$ 
for the fully reconstructed and the \XbJSK channels, respectively. 
The relative yield of the \Xbss, w.r.t. the \Xb yield, is found to be noticeably larger
in the partially reconstructed \Xb channel, compared to the fully reconstructed channels. 
Given the large uncertainties in the observed small \Xbss signal yields, 
this discrepancy is consistent with being a statistical fluctuation. 
The order-of-magnitude larger signals of the \Xbsd decays, 
with respect to the \Xb ground state signals, 
are found to be consistent between all three \Xb reconstruction channels.

The natural width of the \Xbss is too small to be measured with
the present data sample and experimental resolution.
An upper limit on $\Gamma(\Xbss)$ has been obtained through a scan of the 
profiled likelihood, assuming an asymptotic distribution.
The measured upper limit, at 95\% confidence level,
is $\Gamma(\Xbss)<1.9\MeV$, where the systematic uncertainties, 
discussed below, are taken into account.

The local statistical significance of the \Xbss signal is 
evaluated with the likelihood ratio technique, comparing the background-only 
and signal-plus-background hypotheses (with four additional free parameters),
using asymptotic formulas~\cite{Wilks_LL_ratio,Cowan_ssymptotic}.
The resulting significance of the \Xbss signal
varies between 6.2 and 6.7 standard deviations, depending on the
fit model variations used to evaluate the systematic uncertainties.

Several sources of systematic uncertainties in the measured
mass difference \dmXm are considered.
To evaluate the systematic uncertainties related to the choice of the fit model, 
several alternative functions are tested.
Uncertainties related to the choice of the signal model are estimated 
by changing the resolution function from a double-Gaussian function to a single Gaussian function
or a sum of three Gaussian functions.
Two alternative background models are considered:
the threshold function multiplied by an exponential
and the threshold function multiplied by a first-order polynomial.
The largest deviations in the measured mass are 0.01 and 0.04\MeV,
respectively, for the variations of the signal and background models; 
these values are taken as the two corresponding systematic uncertainties.

The RBW function used in the signal modeling includes Blatt--Weisskopf 
barrier factors~\cite{Blatt}, which depend on the radial 
parameter $r$ and on the angular momentum $l$ (spin). 
In the baseline fit, $r=3.5\GeV^{-1}$ and 
$l=1$. The corresponding systematic uncertainties are obtained by varying 
$r$ between 1 and $5\GeV^{-1}$ or by assigning $l=0$.
The $r$ variations have a negligible effect on the results,
while fixing $l=0$ changes the signal shape and induces a 
mass difference variation of 0.01\MeV, taken as the corresponding systematic uncertainty.

To account for a possible difference between the measured and simulated mass resolutions, 
the fits are repeated with resolutions scaled up or down by 1.074, a factor
determined from the comparison of the \Xb resolutions in data and simulation.
The resulting systematic uncertainty of the \Xbss mass difference is 0.02\MeV.

Simulation studies show a shift of 0.07\MeV 
between the generated and reconstructed mass differences; 
this shift is treated as an additional systematic uncertainty in the \dmXm measurement.

The systematic uncertainty reflecting the \dmX fit range 
is evaluated by changing the upper end of the \dmX fit range from its default 100\MeV 
to 80, 120, and 150\MeV. 
The largest mass difference change of 0.02\MeV is taken as the corresponding systematic uncertainty.

A potential bias due to a possible misalignment of the
tracker detectors is evaluated by comparing the results obtained with the data collected in
2016, 2017, and 2018. This is a reasonable
evaluation, given that the inner part of the CMS tracker was replaced
between the 2016 and 2017 data-taking periods. 
The measured mass is found to be insensitive to alignment uncertainties.

The total systematic uncertainty in the measured mass difference \dmXm,
calculated as the sum in quadrature of the partial terms, is 0.09\MeV.

In summary, we report the observation of a new excited beauty strange baryon, 
decaying to \XbPP. The analysis uses proton-proton collision data collected 
by the CMS experiment at \sq{13}\TeV, corresponding 
to an integrated luminosity of 140\fbinv.
The measured mass difference of this state is 
$M(\Xbss) - M(\Xb) - 2\,m^{\mathrm{PDG}}_\pipm = 24.14\pm0.22\stat\pm0.09\syst \MeV$.
The known \Xb mass of $5797.0\pm0.6\MeV$~\cite{PDG} is used to obtain
$ M(\Xbss) = 6100.3 \pm0.2\stat \pm0.1\syst \pm0.6\,(\Xb)\MeV$. 
It is particularly remarkable that if the \Xbss baryon were only 13\MeV heavier,
it would be above the $\Lb\Km$ mass threshold and could decay to this final state.  
The natural width of this resonance is compatible with zero and
a 95\% confidence level upper limit of 1.9\MeV has been determined.

Following analogies with the established excited \PGXc
baryon states~\cite{PDG}, and considering 
several theoretical predictions~\cite{Th:S_P_waves_single_bb, Th:heavy_w_chiral, Th:Xb(6227)_impact}, 
the new \Xbss resonance and its decay sequence are consistent with
the lightest orbitally excited \Xb baryon,
with the light diquark angular momentum $j_{\PQd\PQs}=1$ and $J^P=3/2^-$
(excitation with orbital momentum $L=1$ between the \PQb quark and the $\PQd\PQs$ diquark).
This suggests that it is the beauty analogue of the $\PGXc(2815)$ baryon~\cite{CLEO_Xic(2815)}.
Measuring a natural width of the \Xbss smaller than 1.9\MeV comes as a surprise, 
given the larger values predicted by the theory calculations~\cite{Th:S_P_waves_single_bb, Th:heavy_w_chiral, Th:Xb(6227)_impact},
based on the assumption that the $\Xbstm\to\Xbs\pim$ decay 
proceeds predominantly via $S$-wave ($3/2^-\to3/2^+0^-$).
The observation of this baryon and the measurement of its properties provide
information that should help to distinguish between different theoretical models
used to calculate the properties of the excited \Xbns states.

\begin{acknowledgments}
    We congratulate our colleagues in the CERN accelerator departments for the excellent performance of the LHC and thank the technical and administrative staffs at CERN and at other CMS institutes for their contributions to the success of the CMS effort. In addition, we gratefully acknowledge the computing centers and personnel of the Worldwide LHC Computing Grid and other centers for delivering so effectively the computing infrastructure essential to our analyses. Finally, we acknowledge the enduring support for the construction and operation of the LHC, the CMS detector, and the supporting computing infrastructure provided by the following funding agencies: BMBWF and FWF (Austria); FNRS and FWO (Belgium); CNPq, CAPES, FAPERJ, FAPERGS, and FAPESP (Brazil); MES (Bulgaria); CERN; CAS, MoST, and NSFC (China); COLCIENCIAS (Colombia); MSES and CSF (Croatia); RIF (Cyprus); SENESCYT (Ecuador); MoER, ERC PUT and ERDF (Estonia); Academy of Finland, MEC, and HIP (Finland); CEA and CNRS/IN2P3 (France); BMBF, DFG, and HGF (Germany); GSRT (Greece); NKFIA (Hungary); DAE and DST (India); IPM (Iran); SFI (Ireland); INFN (Italy); MSIP and NRF (Republic of Korea); MES (Latvia); LAS (Lithuania); MOE and UM (Malaysia); BUAP, CINVESTAV, CONACYT, LNS, SEP, and UASLP-FAI (Mexico); MOS (Montenegro); MBIE (New Zealand); PAEC (Pakistan); MSHE and NSC (Poland); FCT (Portugal); JINR (Dubna); MON, RosAtom, RAS, RFBR, and NRC KI (Russia); MESTD (Serbia); SEIDI, CPAN, PCTI, and FEDER (Spain); MOSTR (Sri Lanka); Swiss Funding Agencies (Switzerland); MST (Taipei); ThEPCenter, IPST, STAR, and NSTDA (Thailand); TUBITAK and TAEK (Turkey); NASU (Ukraine); STFC (United Kingdom); DOE and NSF (USA).\end{acknowledgments}

    \hyphenation{Rachada-pisek} Individuals have received support from the Marie-Curie program and the European Research Council and Horizon 2020 Grant, contract Nos.\ 675440, 724704, 752730, and 765710 (European Union); the Leventis Foundation; the Alfred P.\ Sloan Foundation; the Alexander von Humboldt Foundation; the Belgian Federal Science Policy Office; the Fonds pour la Formation \`a la Recherche dans l'Industrie et dans l'Agriculture (FRIA-Belgium); the Agentschap voor Innovatie door Wetenschap en Technologie (IWT-Belgium); the F.R.S.-FNRS and FWO (Belgium) under the ``Excellence of Science -- EOS" -- be.h project n.\ 30820817; the Beijing Municipal Science \& Technology Commission, No. Z191100007219010; the Ministry of Education, Youth and Sports (MEYS) of the Czech Republic; the Deutsche Forschungsgemeinschaft (DFG), under Germany's Excellence Strategy -- EXC 2121 ``Quantum Universe" -- 390833306, and under project number 400140256 - GRK2497; the Lend\"ulet (``Momentum") Program and the J\'anos Bolyai Research Scholarship of the Hungarian Academy of Sciences, the New National Excellence Program \'UNKP, the NKFIA research grants 123842, 123959, 124845, 124850, 125105, 128713, 128786, and 129058 (Hungary); the Council of Science and Industrial Research, India; the HOMING PLUS program of the Foundation for Polish Science, cofinanced from European Union, Regional Development Fund, the Mobility Plus program of the Ministry of Science and Higher Education, the National Science Center (Poland), contracts Harmonia 2014/14/M/ST2/00428, Opus 2014/13/B/ST2/02543, 2014/15/B/ST2/03998, and 2015/19/B/ST2/02861, Sonata-bis 2012/07/E/ST2/01406; the National Priorities Research Program by Qatar National Research Fund; the Ministry of Science and Higher Education, project no. 14.W03.31.0026 (Russia); the Programa Estatal de Fomento de la Investigaci{\'o}n Cient{\'i}fica y T{\'e}cnica de Excelencia Mar\'{\i}a de Maeztu, grant MDM-2015-0509 and the Programa Severo Ochoa del Principado de Asturias; the Thalis and Aristeia programs cofinanced by EU-ESF and the Greek NSRF; the Rachadapisek Sompot Fund for Postdoctoral Fellowship, Chulalongkorn University and the Chulalongkorn Academic into Its 2nd Century Project Advancement Project (Thailand); the Kavli Foundation; the Nvidia Corporation; the SuperMicro Corporation; the Welch Foundation, contract C-1845; and the Weston Havens Foundation (USA).

\bibliography{auto_generated}  \cleardoublepage \appendix\section{The CMS Collaboration \label{app:collab}}\begin{sloppypar}\hyphenpenalty=5000\widowpenalty=500\clubpenalty=5000\vskip\cmsinstskip
\textbf{Yerevan Physics Institute, Yerevan, Armenia}\\*[0pt]
A.M.~Sirunyan$^{\textrm{\dag}}$, A.~Tumasyan
\vskip\cmsinstskip
\textbf{Institut f\"{u}r Hochenergiephysik, Wien, Austria}\\*[0pt]
W.~Adam, J.W.~Andrejkovic, T.~Bergauer, S.~Chatterjee, M.~Dragicevic, A.~Escalante~Del~Valle, R.~Fr\"{u}hwirth\cmsAuthorMark{1}, M.~Jeitler\cmsAuthorMark{1}, N.~Krammer, L.~Lechner, D.~Liko, I.~Mikulec, F.M.~Pitters, J.~Schieck\cmsAuthorMark{1}, R.~Sch\"{o}fbeck, M.~Spanring, S.~Templ, W.~Waltenberger, C.-E.~Wulz\cmsAuthorMark{1}
% ...
\vskip\cmsinstskip
\dag: Deceased\\
1:  Also at Vienna University of Technology, Vienna, Austria\\
% ...
\end{sloppypar}
\end{document}